# Assessing carbon-based anodes for lithium-ion batteries: A universal description of charge-transfer binding

Yuanyue Liu,[1,2] Y. Morris Wang,[3] Boris I. Yakobson,[2] and Brandon C. Wood[1]*

[1]Quantum Simulations Group, Lawrence Livermore National Laboratory, Livermore, California 94550, USA

[2]Department of Materials Science and NanoEngineering, Department of Chemistry, and the Smalley Institute for Nanoscale Science and Technology, Rice University, Houston, Texas 77005, USA

[3]Nanoscale Synthesis & Characterization Laboratory, Lawrence Livermore National Laboratory, Livermore, California 94550, USA

**ABSTRACT:** Many key performance characteristics of carbon-based lithium-ion battery anodes are largely determined by the strength of binding between lithium (Li) and $sp^2$ carbon (C), which can vary significantly with subtle changes in substrate structure, chemistry, and morphology. Here, we use density functional theory calculations to investigate the interactions of Li with a wide variety of $sp^2$ C substrates, including pristine, defective, and strained graphene; planar C clusters; nanotubes; C edges; and multilayer stacks. In almost all cases, we find a universal linear relation between the Li-C binding energy and the work required to fill previously unoccupied electronic states within the substrate. This suggests that Li capacity is predominantly determined by two key factors—namely, intrinsic quantum capacitance limitations and the absolute placement of the Fermi level. This simple descriptor allows for straightforward prediction of the Li-C binding energy and related battery characteristics in candidate C materials based solely on the substrate electronic structure. It further suggests specific guidelines for designing more effective C-based anodes. The method should be broadly applicable to charge-transfer adsorption on planar substrates, and provides a phenomenological connection to established principles in supercapacitor and catalyst design.

The growing demand for energy storage emphasizes the urgent need for higher-performance Li-ion batteries (LIBs). Several key characteristics of LIB performance—namely, reversible capacity, voltage, and energy density—are ultimately determined by the binding between Li and the electrode material.[1-3] Graphite has long been used commercially as a LIB anode, and recently, defective graphene and other $sp^2$ C derivatives have shown promise as high-capacity and high-power anodes.[4-12] However, these seemingly similar substrates exhibit a wide range of Li-C binding energies. For example, pentagon-heptagon pairs are the dominant structural features in both Stone-Wales defects and in certain graphene divacancy complexes, yet theoretical Li binding on the two differs by 0.8 eV.[3] Similar variations are observed for carbon nanotubes (CNTs) with comparable diameters but different chiralities.[13] This in turn contributes to significant variability in the measured voltages and capacities of C-based anodes, ranging from hundreds to thousands of mAh/g. [6, 7, 9, 11, 14-16] Defect incorporation has also demonstrated increases in voltage and capacity,[6, 7, 9, 15] yet the specific defect identities and their role in battery performance merit further exploration. These facts suggest a key factor is missing in the current physical understanding of the underlying Li binding mechanism on C-derived structures, limiting predictive capability.

In this letter, we use plane-wave Density Functional Theory (DFT) calculations to demonstrate how the binding energy of Li on $sp^2$ C-based LIB anode candidates derives from specific features in the intrinsic electronic structure of the substrate, and in most cases can be straightforwardly predicted using a relatively simple descriptor. We further suggest that this same binding descriptor could be generalized to other systems with charge transfer-dominated adsorption behavior. A wide variety of C substrates are considered, including pristine, defective, and strained graphene; graphene-derived molecular clusters; CNTs; C edges modeled by graphene nanoribbons (GNRs); and multilayer graphene. Several point defects are examined: the Stone-Wales (SW) defect, the 5-8-5 ($DV_{585}$) and 555-777 ($DV_{t5t7}$) divacancies, a monovacancy (MV), and single-site substitution by nitrogen ($N_C$) or boron ($B_C$). Calculation details and final adsorption geometries for each substrate can be found in the Supplemental Materials (SM).

The binding energy of a single Li atom is:

$$\varepsilon_{\text{Li-X}} = [E(X) + E(\text{Li-atom}) - E(\text{Li-}X)]/ N_{\text{Li}} \qquad (1)$$

where $E(X)$, $E(\text{Li-atom})$, and $E(\text{Li-}X)$ are the energies of the Li-free substrate $X$, an isolated Li atom, and the Li-adsorbed substrate $X$, respectively. $N_{\text{Li}}$ is the number of adsorbed Li atoms in the supercell. We report the values with respect to the cohesive energy of bulk Li ($\varepsilon_{\text{Li-Li}}$) according to $\Delta\varepsilon_{\text{Li-X}} \equiv \varepsilon_{\text{Li-Li}} - \varepsilon_{\text{Li-X}}$; lower values represent stronger binding.

Upon dilute Li adsorption (concentrations below $LiC_{18}$, where $LiC_n$ represents a Li : C ratio of 1 : n), the Dirac cone near the Fermi level ($\varepsilon_f$) of pristine graphene retains its shape, while $\varepsilon_f$ is shifted to a higher energy, as shown in Fig. 1a. The number of occupied states above the Dirac point (DP) equals the number of Li atoms, indicating complete ionization of Li via charge transfer to the substrate. The states occupied upon Li-to-C electron transfer have completely delocalized π character (left-hand panel, Fig. 1b). We refer to this behavior as "states-filling", as it describes a rigid occupation shift against a backdrop of otherwise unchanged π states in the vicinity of $\varepsilon_f$. In this respect, Li on graphene appears to behave similarly to an electronic dopant in the energy window near $\varepsilon_f$. This response differs from that of many transition metal adatoms, which tend to create new states within the Dirac cone.[17, 18] Nevertheless, we emphasize that the effect of Li is not that of *pure* electronic doping, since the potential from the adsorbed ion also alters the character of the deeper valence states; this can be seen in the center panel of Fig. 1b, in which valence charge density has accumulated near the ionized $Li^+$ adsorbate.

The states-filling behavior of pristine graphene is largely retained for almost all of the other substrates we tested, provided binding occurs on the basal plane (edge binding is discussed later). In each case, Li acts as a dopant near $\varepsilon_f$, donating its electron to previously unoccupied π C states without introducing additional bands. As an example, Fig. 1c shows the band structure of a $DV_{t5t7}$ point defect undergoing a mostly rigid shift upon Li binding. Other tested point defects exhibit analogous behavior, despite their strongly dissimilar electronic structures (band structures for each can be found in the SM).

If we consider only the electronic doping character (i.e., rigid band shift) near $\varepsilon_f$, then states-filling behavior suggests that $\varepsilon_{Li-X}$ should correlate with the work required to fill empty C states with the Li-donated excess electron. With all energies referenced to the vacuum level, this work is defined as (per Li):

$$W_{\text{filling}} = \int_{\varepsilon_{LUS}}^{\varepsilon'} \varepsilon D(\varepsilon)/N_{Li} d\varepsilon \quad (2)$$

where $\varepsilon$ is the Kohn-Sham (KS) energy, $D(\varepsilon)$ is the density of states (DOS) of the Li-free C supercell, and $\varepsilon'$ satisfies the charge-conservation criterion:

$$\int_{\varepsilon_{LUS}}^{\varepsilon'} D(\varepsilon)/N_{Li} d\varepsilon = 1 \quad (3)$$

Here, LUS is the lowest unoccupied state: the Fermi level ($\varepsilon_f$) for a metal, the conduction-band minimum ($\varepsilon_{CBM}$) for a non-metal, or the LUMO level for a finite system. We refer to Eqns. 2 and 3 as the states-filling model (SFM). There are two cases of Equation 2 that deserve special consideration: (1) on a finite cluster, $\varepsilon$ is discrete, and $W_{\text{filling}}$ becomes the LUS (LUMO); (2) in the infinitely dilute adsorption limit, $D(\varepsilon)/N_{Li}$ diverges, and $W_{\text{filling}}$ again converges to the LUS ($\varepsilon_f$ or $\varepsilon_{CBM}$). Note that $W_{\text{filling}}$ implicitly depends on two factors: the C electronic structure and the Li concentration.

Examination of the dependence of $\Delta\varepsilon_{Li-X}$ on $W_{\text{filling}}$ for dilute Li adsorption on a wide variety of $sp^2$ C forms shows that not only are the quantities indeed positively correlated, but that the relation is linear for each class of substrate modification (Fig. 2). The simplicity of the result is surprising, since the SFM deliberately ignores all perturbations to the deeper valence states. Fig. 2a shows the linearity with varying Li concentrations on pristine graphene up to $LiC_{72}$ (dense adsorption is addressed later). Increasing the concentration requires more high-energy states to be filled, which raises both $W_{\text{filling}}$ and $\Delta\varepsilon_{Li-graphene}$. Note that there is a concentration dependence of $W_{\text{filling}}$ even at very dilute adsorptions, a consequence of the delocalized nature of the newly filled π states (Fig. 1b). Fig. 2b shows the effect of isotropic tensile strain at fixed Li concentration ($LiC_{72}$) on graphene. $W_{\text{filling}}$ decreases with increased strain, and Li binding is stabilized. In Fig. 2c, Li is adsorbed on graphene with various point defects (~ $LiC_{72}$), which have very different electronic structures and hence a wide range of $W_{\text{filling}}$ values. Here we test two scenarios, one with Li placed in a region away from the defect and another with Li placed directly on the defect site (see SM). Either way, the linearity is manifest, deviating only slightly for direct adsorption on $DV_{t5t7}$ and $B_C$.

The slope at the defect sites is steeper, reflecting additional changes to the low-energy states (confirmed by visualization of the electron accumulation). Fig. 2d shows the dependence of $\Delta\varepsilon_{\text{Li-cluster}}$ on the size of a finite graphene-like cluster. Smaller clusters have larger band gaps, which result in higher $W_{\text{filling}}$, and consequently, higher $\Delta\varepsilon_{\text{Li-cluster}}$. In Fig. 2e, Li is adsorbed on several chiralities of CNTs (~ $LiC_{600}$) with similar diameters (9.0-9.8 Å). Metallic tubes have the lowest $W_{\text{filling}}$ and the strongest binding.

With all data viewed globally, the positive correlation between $\Delta\varepsilon_{\text{Li-X}}$ and $W_{\text{filling}}$ is clear (Fig. 2f). However, each type of modification has a unique slope and intercept within its individual linear relation. If rigid band shifts were solely responsible for the differences in Li-C binding, then one should always expect a slope of unity, yet this is not generally the case. In KS DFT, the total energy is:

$$E = \int \varepsilon D(\varepsilon) d\varepsilon - \frac{e^2}{2} \int dr \int dr' \frac{\rho(r)\rho(r')}{|r-r'|} + E_{\text{ion-ion}} \qquad (4)$$

where successive terms represent the total energy of the occupied KS eigenstates, the Hartree energy, and the ion-ion Coulomb energy. In the SFM, $W_{\text{filling}}$ directly accounts only for energy changes in the states above $\varepsilon_{\text{LUS}}$ under the rigid-band approximation, which contribute to the first term in Equation (4). Other possible contributions to $\varepsilon_{\text{Li-X}}$ that are not included in the SFM include: (1) deviations from the rigid-band approximation or changes in eigenstates below $\varepsilon_{\text{LUS}}$, (2) changes in the Hartree energy, and (3) changes in $E_{\text{ion-ion}}$. Notably, the observed universal linearity between $\varepsilon_{\text{Li-X}}$ and $W_{\text{filling}}$ leads to the nontrivial conclusion that *all collective remaining contributions to $\varepsilon_{\text{Li-X}}$ must also depend linearly on $W_{\text{filling}}$*. We suspect that this dependence derives in part from two factors contained in $W_{\text{filling}}$ that also determine the screening of the adsorbate-induced electric field within the substrate:[19-21] the adsorbate concentration and the DOS at $\varepsilon_f$. As the concentration increases and the DOS decreases (i.e., fewer available states and generally larger $W_{\text{filling}}$), screening becomes poorer and the electronic density becomes more inhomogeneous, impacting the effective Hartree potential. Within this interpretation, our observed linear relation is consistent with recent calculations by Santos and Kaxiras,[22] who demonstrated a similar linear dependence between the in-plane electric susceptibility of graphene ribbons and the number of available atoms (i.e., states) across which charge may be redistributed.

The success of the SFM straightforwardly explains the observed diversity in $\varepsilon_{\text{Li-X}}$ values across substrates. For instance, the SW and $DV_{t5t7}$ defects, both comprised of pentagon-heptagon pairs, have very different electronic structures: the $DV_{t5t7}$ defect has a lower $\varepsilon_f$ (below the DP) due to its missing C atoms/electrons, and a higher DOS near $\varepsilon_f$ (compare Figs. 1c and S1). As a result, $DV_{t5t7}$ shows lower $W_{\text{filling}}$ and stronger Li binding.[3] Similarly, the lower $W_{\text{filling}}$ of metallic CNTs with respect to semiconducting CNTs explains the stronger Li binding to the former.[13] In addition, substitutional $B_C$ and $N_C$ have similar DOS at $\varepsilon_f$,[23] yet the former has stronger $\varepsilon_{\text{Li-X}}$ due to its lower $\varepsilon_f$;[3] this is borne out in experiments demonstrating higher capacity for B treatment than N treatment.[7]

The DOS dependence of $W_{filling}$ in Equation (2) suggests that a key limitation of Li capacity in graphene derivatives lies in how easily excess electrons can be absorbed. This is closely related to the quantum capacitance $C_q(V) = e^2 D(\varepsilon)$, the integral of which gives the potential-dependent ($V$) charge storage capacity.[23, 24] A high $C_q(V)$ near $\varepsilon_f$ therefore correlates with stronger binding. Accordingly, the intrinsically poor quantum capacitance of graphene becomes a vital consideration in the design of higher-capacity LIB anodes, much as it does in the design of C-based supercapacitors[23, 25] and field-effect transistors.[26, 27]

According to Equation (2), $W_{filling}$ also depends on the vacuum-referenced $\varepsilon_{LUS}$. In this regard, the SFM is a charge-transfer-binding analogue to the "$d$-band center" theory in transition-metal catalysis, which connects a higher metal $d$-band center to easier filling of antibonding states, and hence to stronger binding.[28] It also justifies the observed dependence of surface molecular dissociation barriers (related to binding strength) on the catalyst work function, which converges to the vacuum-referenced $\varepsilon_f$ ($\varepsilon_{LUS}$) for high-DOS metals.[29]

Significantly, the SFM suggests simple guidelines for designing effective $sp^2$ C-based anodes, since a low $\varepsilon_{LUS}$ and high $C_q(V)$ will lead to stronger Li binding and typically higher Li capacity. Accordingly, electron-withdrawing groups and $p$-type dopants are good candidates, which explains why materials such as $BC_3$ have high theoretical capacities.[3] Point defects[3] and curvature may also improve capacity, since they tend to elevate $C_q(V)$ near $\varepsilon_f$.[23] This probably contributes to experimentally observed voltage and capacity increases upon defect incorporation,[6, 7, 9, 15] which contrasts with the low Li adsorption limits found for pristine graphene.[30]

Although we have specifically developed the SFM to explain the physical principles underlying Li binding on the $\pi$ manifold of $sp^2$ C, we emphasize that it should be generalizable to other systems and applications where charge transfer dominates the adsorption behavior. Nevertheless, there are some important conditions for its application. First, it contains no information about the site dependence of the binding energy, since it is based on the total DOS. Accordingly, it is best applied to systems where such sensitivity is low, such as when newly occupied states are delocalized.

Second, the SFM assumes charge transfer to the substrate is complete. As such, it fails at very high Li concentrations on low-DOS substrates, where the energetic cost for excess charge storage is large enough that charge transfers back to the Li as free-electron states (Fig. 3a). On pristine graphene, this occurs at concentrations beyond ~ $LiC_8$ ($W_{filling}$ ~ -3.0 eV), lowering $\Delta\varepsilon_{Li-graphene}$ and leading to deviations from ideal states-filling behavior (Fig. 3b). This places an absolute limit on the $W_{filling}$ for which the SFM is expected to hold; once free-electron Li-derived states are introduced, adsorption can no longer be treated as dilute. Nevertheless, experimentally realizable reversible capacities of less-disordered C-based materials often fall well below this dilute threshold. [9]

Third, the SFM relies on band rigidity near $\varepsilon_{LUS}$, and therefore does not apply when bands/states are created in this region upon adsorption. One example is σ-binding of Li to graphene edges, where the Li electron localizes at the edge atoms and creates a new flat band (Fig. 3c).

Fourth, the SFM assumes that the within a modification class, the presence of the adsorbate perturbs deeper electronic states similarly for every value of $W_{filling}$. This prevents direct comparisons between surface-adsorbed graphene and intercalated graphite, since the latter exhibits a qualitatively different π-electron density distribution due to overlap of the electrostatic potential wells of the individual stacked Li-adsorbed graphene sheets (Fig. 3d).[31] As a result, graphite binds Li stronger than graphene by 0.7 eV (at $LiC_6$) [3, 31] yet has a very similar $W_{filling}$.

An added advantage of the SFM is that binding properties can be quickly predicted based only on the substrate electronic structure, which is useful for rapid primary screening. For instance, we can easily estimate the Li capacity of a candidate C-based anode, which is the critical concentration $c$ satisfying:

$$\varepsilon_{Li\text{-}X}(c) + c\frac{d\varepsilon_{Li\text{-}X}(c)}{dc} = 0 \qquad (5)$$

Solving Equation 5 requires the continuous $\varepsilon_{Li\text{-}X}(c)$ function, which necessitates a large supercell calculation for each discretized value of $c$. If we apply the SFM, only two concentrations are needed to obtain the linear $\varepsilon_{Li\text{-}X}(W_{filling})$ equation for the chosen class of surface modification. By extracting $W_{filling}(c)$ from the continuous DOS of the unlithiated primitive cell, $\varepsilon_{Li\text{-}X}(c)$ and the corresponding Li capacity can then be straightforwardly determined. Capacity values obtained in this way show good agreement with explicit calculations of $\varepsilon_{Li\text{-}X}(c)$ (see SM). A second example in the SM shows how the dependence of $\varepsilon_{Li\text{-}X}$ on tensile strain can be easily computed at any given Li concentration.

In summary, we propose a simple descriptor that captures the essential physics of charge transfer-dominated binding on planar carbon, based on the work required to fill up the rigid electronic states of the substrate. Applied to C-based LIB anodes, our model explains the physical origin of the wide range of Li-C binding energies reported in the literature, and suggests a link to the significant variability in the reported performance of graphene-derived anodes. It also provides guidelines for engineering more effective anodes; these predictions are consistent with experimentally demonstrated improvements via substrate modification. By drawing upon similar considerations to those used in catalyst and supercapacitor electrode design, the descriptor straightforwardly connects anode performance to intrinsic electronic structure and establishes the broader role of the latter in interfacial electrochemical systems.

We thank T. Ogitsu, T. W. Heo, Y. An, J. Ye, M. Tang, and V. Artyukhov for valuable discussions. Funding was provided by LLNL LDRD Grant 12-ERD-053, with computing support from the LLNL Institutional Computing Grand Challenge program. This work was performed under the auspices of the U.S. Department of Energy by LLNL under Contract DE-AC52-07NA27344.

———

*brandonwood@llnl.gov

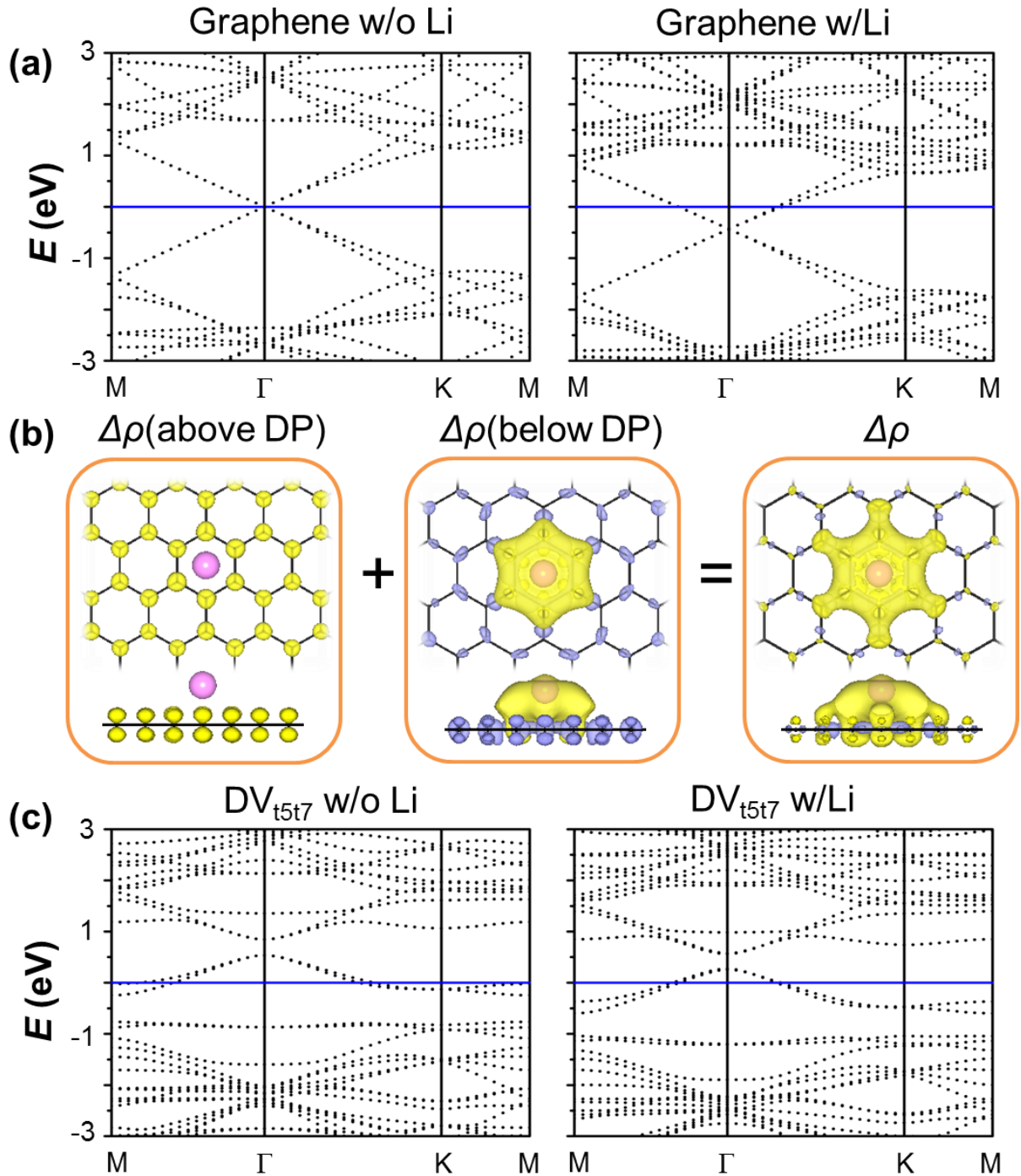

**Figure 1**. (a) Band structure of graphene (6x6 cell) without (left) and with (right) Li. The Fermi level (blue line) is set to zero. (b) Charge density difference between Li-free and Li-adsorbed graphene for states above (left) and below (middle) the Dirac point, and for all states (right). Electron accumulation (depletion) upon Li adsorption (purple) is indicated by the yellow (blue) isosurface of $10^{-3}$/Bohr$^3$. (c) Band structure of a $DV_{t5t7}$ defect (6x6 cell) without (left) and with (right) Li.

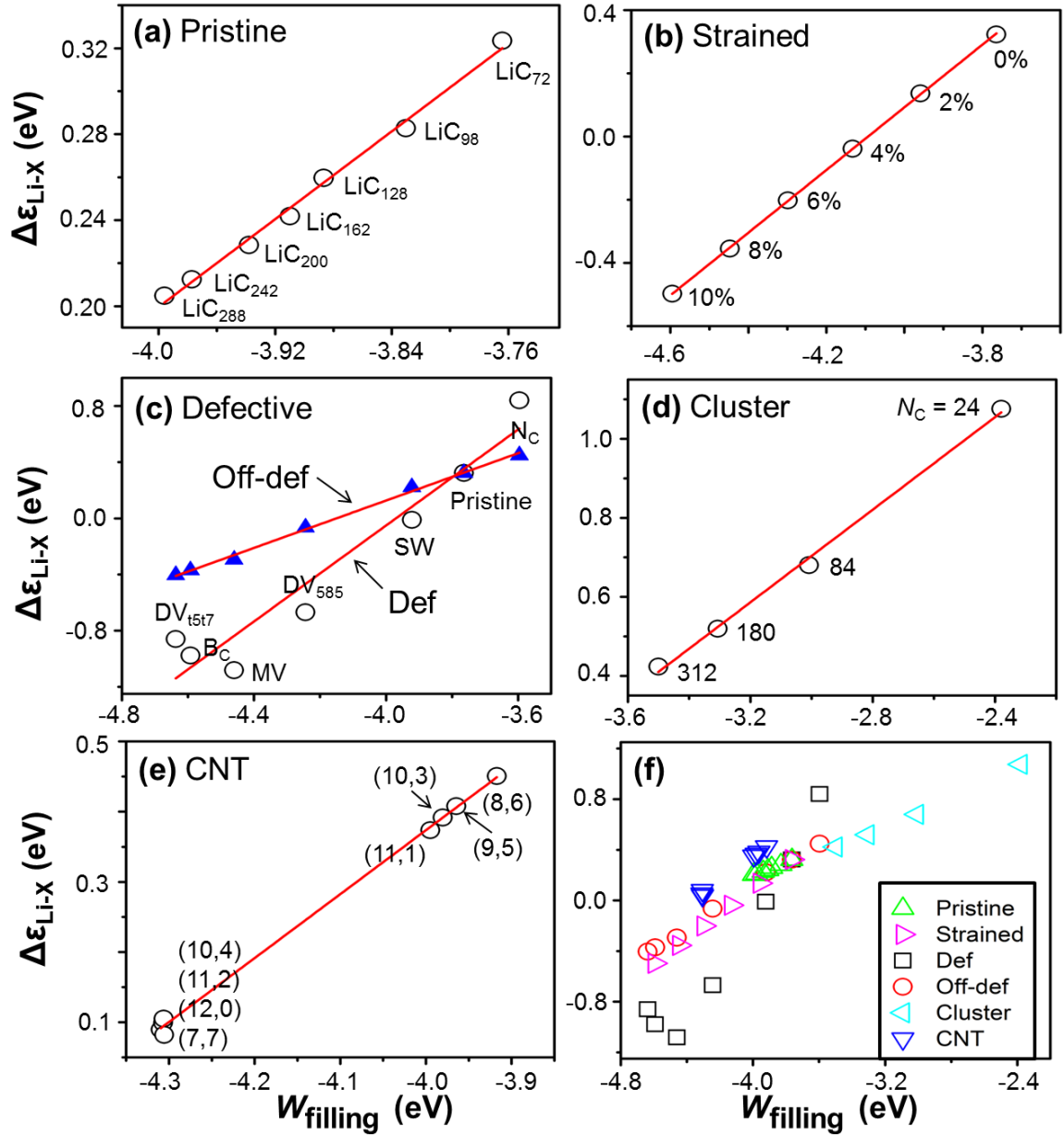

**Figure 2**. Linear dependence of $\Delta\varepsilon_{\text{Li-X}}$ on $W_{\text{filling}}$ for: (a) pristine graphene with different Li concentrations; (b) graphene under varying isotropic tensile strain, based on the percent increase in the lattice parameter; (c) defective graphene, where black circles represent adsorption directly at a defect site (Def), and blue triangles at an off-defect region (Off-def); (d) different-sized hexagonal graphene clusters with $N_C$ C atoms and Li adsorption at the center; (e) CNTs of similar diameter (9.0-9.8 Å) but different chiralities (~ $LiC_{600}$), with top-right (bottom-left) points representing semiconducting (metallic) CNTs; (f) all tested substrates. Red lines are linear fits; fitting parameters are given in the SM.

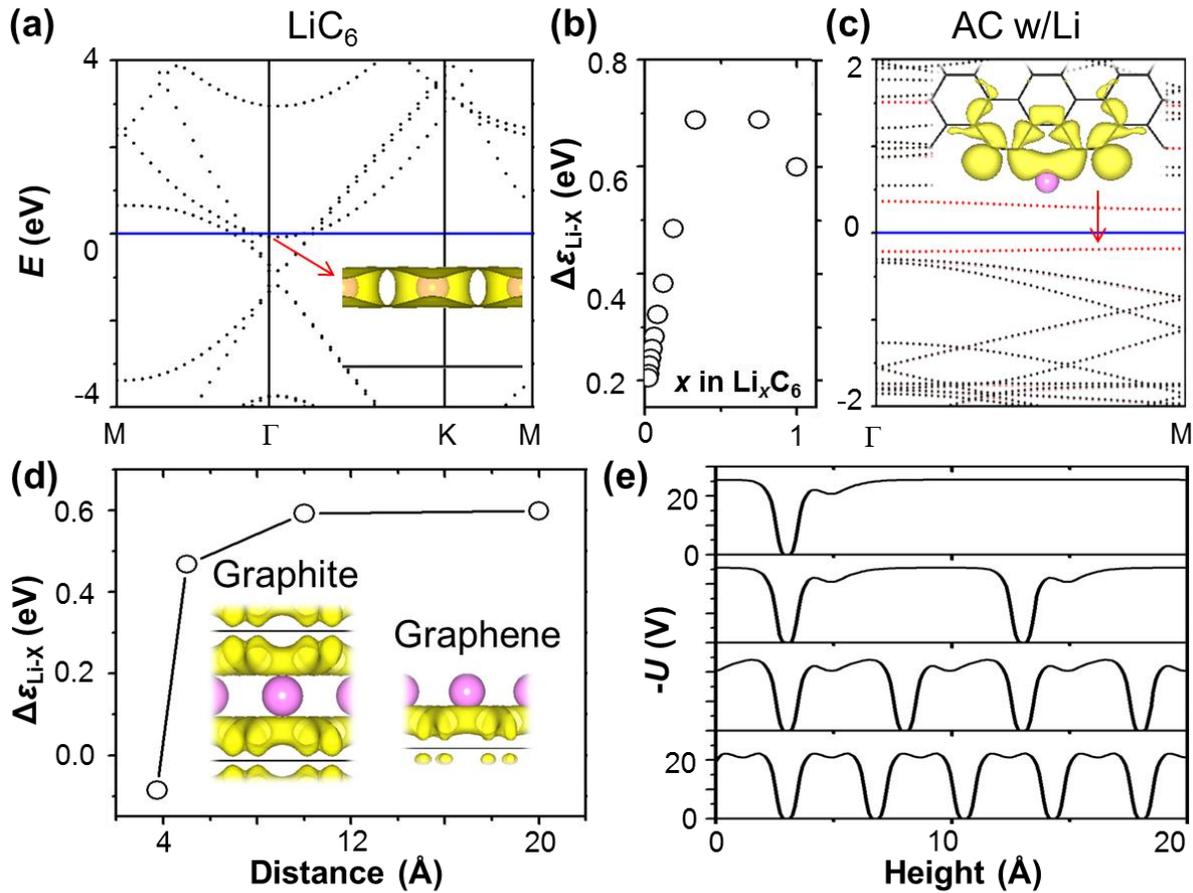

**Figure 3**. (a) Band structure of graphene with dense Li concentration (LiC$_6$). (b) Concentration dependence of $\Delta\varepsilon_{\text{Li-graphene}}$, showing the breakdown of linear dependence at high Li loading. (c) Band structure of a GNR with Li adsorbed at an armchair (AC) edge (black/red are spin up/down). Insets for (a) and (c) show charge density contributions from the states marked by arrows. (d) $\Delta\varepsilon_{\text{Li-graphene}}$ as a function of separation between periodically stacked LiC$_6$ layers, with the corresponding evolution of the Li-induced electron accumulation shown. (e) Electrostatic potential (-$U$; minimum set to zero) normal to stacked LiC$_6$ layers at the separations in (d), decreasing from top to bottom.